\documentstyle{mn}
\title[SuperCOSMOS Sky Survey III: Astrometry]{The SuperCOSMOS Sky Survey. Paper
III: \\ Astrometry}
\author[N.C.\ Hambly et al.]{N.C.\ Hambly$^{\rm 1}$, A.C.\ Davenhall$^{\rm 2}$, 
M.J.\ Irwin$^{\rm 3}$ and H.T.\ MacGillivray$^{\rm 1}$\\
$^1$Wide Field Astronomy Unit, Institute for Astronomy, University of Edinburgh, Blackford Hill, Edinburgh, EH9~3HJ\\
$^2$Institute for Astronomy, University of Edinburgh, Blackford Hill, Edinburgh, EH9~3HJ\\
$^3$Cambridge Astronomy Survey Unit, Institute of Astronomy, Madingley Road,
Cambridge, CB3~0HA\\
}

\date{Accepted ---. 
      Received ---;
      in original form ---}

\begin{document}

\maketitle

\begin{abstract}
In this, the third in a series of three papers concerning the SuperCOSMOS
Sky Survey, we describe the astrometric properties of the database.
We describe the algorithms employed in the derivation of the
astrometric parameters of the data, and demonstrate their accuracies by
comparison with external datasets using the first release of data,
the South Galactic Cap survey. We show that the
celestial co-ordinates, which are tied to the International
Celestial Reference Frame via the Tycho--2 reference catalogue,
are accurate to better than $\pm$~0.2~arcsec at 
J,R~$\sim$~19,18 rising to $\pm$~0.3~arcsec at J,R~$\sim$~22,21 with 
positional dependent systematic effects from bright to faint magnitudes at the 
$\sim$~0.1~arcsec level. 
The proper motion measurements are shown to be accurate
to typically $\pm$~10~mas~yr$^{-1}$ at J,R~$\sim$~19,18 rising to 
$\pm$~50~mas~yr$^{-1}$ at J,R~$\sim$~22,21 and are tied to zero using the
extragalactic reference frame. We show that the zeropoint errors in the
proper motions are $\leq1$~mas~yr$^{-1}$  for R~$>17$ and are no larger than
$\sim10$~mas~yr$^{-1}$ for R~$<17$~mas~yr$^{-1}$. 
\end{abstract}

\section{Introduction}
\label{intro}

In Paper~{\sc I} of this series (Hambly et al.~2001a) 
we describe the SuperCOSMOS Sky Survey (hereafter SSS)\footnote{database 
available online at \verb+http://www-wfau.roe.ac.uk/sss+}. 
This ambitious project ultimately aims to digitise
the entire sky atlas Schmidt plate collections in three colours (BRI), one
colour (R) at two epochs. Paper~{\sc II} in this series
(Hambly, Irwin \& MacGillivray~2001b)
describes the image detection, parameterisation, classification and photometric
calibration techniques for the survey. In this, the third paper of the series,
we describe in detail the calibration of the astrometric parameters contained
within the survey data. Paper~{\sc I} is intended as a user guide to the SSS
while Papers~{\sc II} and~{\sc III} provide more technical details concerning
the derivation and calibration of the object parameters. 
The first release of data consisted of $\sim5000$ 
square degrees of the southern sky at high Galactic latitude ($|b|>60^{\circ}$), and is known as the South Galactic Cap survey. Although these three
papers make explicit reference to SGC data, all details are 
generally applicable to SSS data at Galactic latitudes $|b|\geq30^{\circ}$.
At low Galactic latitudes, particularly towards the Galactic centre, crowding
degrades astrometric and photometric performance.

The photographic plate material used for the SSS survey consist of sky--limited
Schmidt photographic glass plate and film originals, or glass copies of glass
originals, taken with the UK, ESO and Palomar Oschin Schmidt Telescopes -- for
more details see Paper~{\sc I}, Morgan et al.~(1992) and
references therein. Leaving aside such details as object detection and 
parameterisation (see Paper~{\sc II}) the solution to the problem of assigning
celestial co-ordinates to each object requires comparison with a reference 
catalogue and the application of a plate model. Schmidt plate astrometry has
a long history (see Section~\ref{astsol} for some references) involving
reference catalogues of increasing density and precision along with 
increasingly sophisticated plate models -- for a concise review, see Morrison,
Smart \& Taff~(1998). These authors
question the need for a conventional plate model at
all and advocate a novel technique making use of an `infinitely overlapping
circles filter'. However, we note that they concede that,
for large--scale survey work at least, a global linear plate model plus
non--linear correction via an empirical plate `mask' is appropriate. It is
just such a method that we describe in Section~\ref{astsol}. In 
Section~\ref{results}, we proceed to test the astrometric solutions
against CCD drift scan data encompassing plate boundaries -- ie.~a 
test of the true external errors, both random and systematic, as a function of
magnitude and plate position. Section~\ref{results} also discusses our results
in comparison to extragalactic objects defining the International Celestial 
Reference Frame (Ma et al.~1998).

In addition to providing celestial co-ordinates, the SSS survey catalogues
include
proper motions. In the southern hemisphere, all objects paired between the 
SERC--J/EJ (first epoch, hereafter B$_{\rm J}$) and SERC--ER/AAO--R (second 
epoch, hereafter R) plates have a proper motion
measurement (see Section~\ref{pmchoice}). The
methodology for determining the proper motions closely follows that described
in Evans \& Irwin~(1995; hereafter EI95) and is detailed in Section~\ref{pms}.
Small changes made necessary by the plate material used are also described.
Section~\ref{results} contains a comparison between our derived proper motion
measurements against an external dataset, and we analyse in some detail the
relative contribution of random and systematic errors on these
proper motions as a function of magnitude.

As we will describe in Section~\ref{astsol} the celestial co-ordinates are 
tied to the Hipparcos reference frame via the Tycho--2 catalogue. As stated in
the introduction to the Hipparcos and Tycho Catalogues (ESA~1997) this frame
is practically identical to the International Celestial Reference System,
and we demonstrate the level of any residual zeropoint errors with respect
to the ICRF in Section~\ref{results}.
Furthermore, our proper motion zeropoint is taken from
extragalactic objects, and is therefore also tied to an inertial reference
frame. The advantages of such a proper motion reference frame are discussed
extensively with reference to the Lick Observatory Northern Proper Motion 
programme (NPM -- Klemola, Jones \& Hanson~1987) and are clearly demonstrated
by recent results from the corresponding southern hemisphere survey, the
Yale Observatory/San Juan Southern Proper Motion program (SPM --  Platais et
al.~1998; M\'{e}ndez et al.~1999).

\section{Methodology}
\label{methods}

All photographic material for the SSS survey is measured on SuperCOSMOS, a 
fast high precision microdensitometer. Aspects of the system enabling highly
accurate positional measurements are described in Hambly et al.~(1998) and
will not be repeated here, except to state that it was demonstrated that the
machine has the capability to centroid isolated, well--exposed stellar images on 
modern fine--grained emulsions at a precision of $\sim0.5\mu$m (or 33mas at
the plate scale of the sky survey Schmidt plates).

\subsection{Global astrometric plate solutions}
\label{astsol}

Converting the positions of stars and other objects measured on a 
photographic plate into celestial coordinates is a classical problem
in photographic astronomy and various techniques are available to solve
it.  All the techniques are based on having a grid of reference stars,
with known celestial coordinates, distributed over the plate.  A
transformation is then defined between the measured $xy$ positions
and the celestial coordinates of the reference stars. 
It is the details of calculating the transformation which may vary
between different techniques.

The SuperCOSMOS global astrometric software (hereafter XYTORADEC)
uses an algorithm based on the {\sc Starlink} program ASTROM (Wallace~1998).
The code implementing the algorithm is partly borrowed from ASTROM and
partly consists of calls to routines in SLALIB (Wallace~1999), the 
{\sc Starlink}
Positional Astronomy Library.  ASTROM (and hence XYTORADEC) use a modified
version of the standard, traditional technique for converting plate
positions to coordinates.  In this technique the catalogue coordinates
of the reference stars are converted to tangent plane coordinates,
adjusted for geometrical distortion (caused by the telescope optics 
and mechanical deformation of the plates during exposure) and a
linear six-coefficient least squares fit is made between the adjusted
standard coordinates and the measured positions.  Several authors have
described this technique, eg.~K\"{o}nig~(1962), Taff~(1981) and Green~(1985).
In this method a simple, single, global
fit is made for all the reference stars imaged on the plate.  The
technique used by ASTROM and XYTORADEC differs from the traditional
method in that the plate centre is allowed to vary 
(that is, included in the fit).  An outline of the method is as follows:

First, the point on the celestial sphere
at which the telescope was pointing when the plate was exposed is
established. Reference star coordinates are converted to apparent (that is,
observed) coordinates in the desired system. Standard (tangent plane) coordinates, $\xi\eta$, are computed
for the reference stars about the
tangent point. The standard coordinates are then adjusted for the 
geometrical distortion where each image centroid is
adjusted radially.  The adjustment consists of separately
multiplying each $\xi$ and $\eta$ by the factor
$r/\tan r$ where $r=(\xi^2+\eta^2)^{1/2}$ (see Murray~1983 page 196).

\begin{figure*}
\begin{center}
\setlength{\unitlength}{1mm}
\begin{picture}(150,110)
\includegraphics{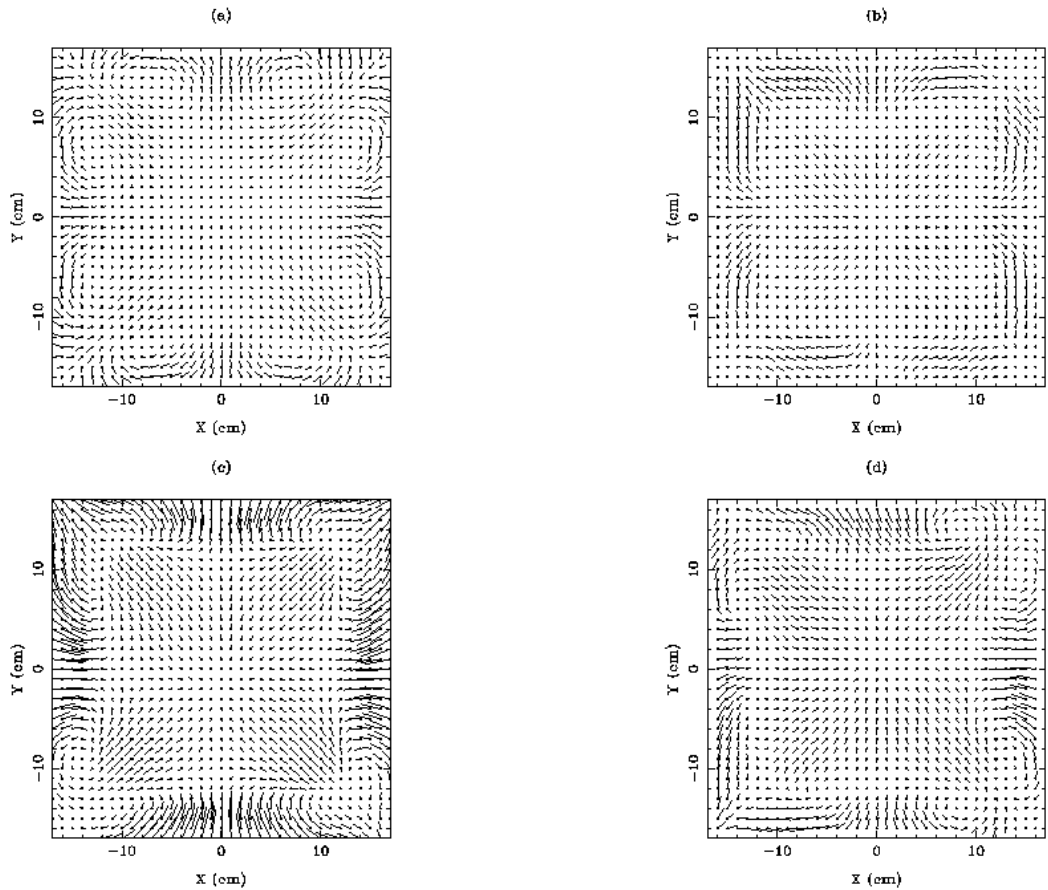}
\end{picture}
\end{center}
\caption[]{Mechanical deformation (or `swirl')
patterns for (a) UK Schmidt SERC--J/EJ plates, (b)
ESO Schmidt R plates, (c) Palomar Schmidt first epoch E plates and 
(d) Palomar Schmidt second epoch R plates. In each case, the
length of the vectors is scaled such that one tick--mark corresponds to one
arcsecond.}
\label{swirls}
\end{figure*}

After applying the geometric distortion a small systematic
   distortion remains between the adjusted standard coordinates and
   positions measured on Schmidt plates.  These residuals are the
   well-known `swirling patterns' seen in the residuals of
   linear fits made to Schmidt data (eg. Taff et al.~1992).
   A set of corrections at a grid of points over the plate are
   prepared beforehand. 
A grid size of 1~cm on the plate ($\sim10$~arcmin) was chosen since this
provides sufficient resolution to accurately map out the $\sim30$~arcmin--scale
non--linear distortion expected.
   Different grids are used for plates originating from
   different Schmidt telescopes and in different wavebands.  These grids are
   prepared by averaging the residuals over numerous plates, using this
fitting procedure (but of course without these corrections).
Figure~\ref{swirls} shows four examples of the distortion patterns from
the three Schmidt telescopes providing data for sky surveys. For the
UK Schmidt J~survey, 
the data are averaged over 200 plates containing 193866 individual
standard star residuals. The ESO Schmidt data are from 142 plates/107346 stars
while for the Palomar Schmidt first epoch~E and second epoch R~surveys
the data are from 51 plates/47101 stars and 9 plates/20814 stars respectively.
In all cases, the reference catalogue used was the Tycho--2 (H\o g et al.~2000).
The largest non--linear effects are seen 
for the Palomar POSS--I~E plates. This has been seen previously, and is
probably a result of the design of the original plate holders used for that
survey (eg.~Irwin~1994). Note that the redesign of these plate holders has
resulted in much smaller non--linearities for the second epoch 
survey -- compare Figures~\ref{swirls}(c) and~(d).
   For each reference star or programme object the appropriate
   correction is derived by interpolating the grid at the position
   of the star or object.  Simple bilinear interpolation is used
   (eg. Press et al.~1986).  The corrections
   are then simply added to the standard coordinates.

  The standard coordinates, adjusted for mechanical and optical distortion,
   are then fit to the measured positions.  
For a complete description of XYTORADEC see Davenhall~(2000).
Section~\ref{results} illustrates typical results from the above procedure.

\subsubsection{Reference catalogue.}

The SGC survey data originally made publicly available on the World Wide Web
made use of the ACT reference catalogue (Urban et al.~1998). Since mid--2000,
the Tycho--2 catalogue has been available (H\o g et al.~2000),
and all SGC astrometry was
re-reduced with respect to this catalogue. SSS data are now routinely reduced
with respect to the Tycho--2 reference catalogue.
This compilation of 2.5 million stars reaches magnitudes 
V~$\sim12.5$ (90\% complete to V~$\sim11.5$), and has 
positions at mean observational epoch of $\sim$~J1991.5 from the 
Tycho catalogue, but includes proper motions via a combination of
reprocessed Tycho data and first epoch positions from the 
Astrographic Catalogue and 143 other ground--based catalogues
(see references in H\o g et al.). The astrometric
standard stars provided by Tycho--2 are heavily saturated on sky--limited
Schmidt plates, and as we show in Section~\ref{results} there is a demonstrable
magnitude systematic between bright and faint stars at the level of up to
$\sim0.3$~arcsec at field edges.
Future projects, both ground--based (for example, the UCAC
project -- see Zacharias et al.~2000) and space--based, will
provide higher density catalogues reaching fainter magnitudes and will
bridge the magnitude range ($10.0<{\rm V}<15.0$) required to
eliminate such systematic errors from the plate solutions
(e.g.\ Irwin et al~1998). In this way, the
full--blown {\em internal} precision obtainable from centroiding on Schmidt
plates will ultimately limit the external accuracy of the astrometry. We note
here that the SSS survey database is organised in such a way as to make 
rereduction of the plate astrometry easy to achieve, and when higher density
catalogues become available this will indeed be done.

\subsubsection{FITS World Co-ordinate System for images extracted from the SSS
database.}

The availability of global astrometric plate solutions for the survey data 
enables any extracted FITS image (see Paper~{\sc I}) 
to contain FITS World Co-ordinate
System (WCS) header information. As yet, there is no formally accepted standard
method and keyword set for implementing the WCS. However, Greisen \& 
Calabretta~(1999) have proposed a standard, and we have followed this in our
database access software. The WCS is defined locally for each extracted image
by fitting a linear plate model between tangent plane projected celestial
co-ordinates and $xy$ in pixel units. Large--scale, non--linear distortions
are not relevant to small scales of tens of arcminutes. We simply specify
the WCS using the \verb+RA---TAN+ and \verb+DEC--TAN+ prescription along with
the \verb+CD+ rotation matrix elements (see Greisen \& Calabretta~1999 for
more details). This set is compatible with, for example, the 
{\sc Starlink} software collection (eg.~image display utilities such as
GAIA/SkyCAT, Draper~1999). Note also that previously defined keywords sets
are also written to the FITS headers for backwards compatibility.

\subsubsection{R survey data on film}

At the time of writing, it seems likely that the AAO--R survey will be
completed on Kodak 4415 `Tech--Pan' film. For example, a small number 
($\sim25$) of SGC fields have film originals measured for the database.
Due to its finer emulsion granularity, this medium is superior to IIIaF in
all aspects of astronomical performance 
(eg.\ Parker \& Malin~1999; Parker et al.~2001) with the
exception of large scale astrometric stability. The `swirl' patterns seen
for glass plates (eg.~Figure~\ref{swirls}) do not repeat from film to film,
but systematics at similar scales and of a similar amplitude are nonetheless
present. Such fields have individual systematic correction via the
corresponding J~plate data. The procedure is simply to map out all 
plate--to--plate errors as a function of position between the R~film and its
corresponding SERC--J/EJ~plate in a given field using the J~plate to define
the `swirl' correction mask.

\subsection{Proper motions}
\label{pms}

\subsubsection{Choice of plate pairs and pairing criterion}
\label{pmchoice}

The plates available for the SSS provide several potential plate
pair combinations for the determination of proper motions. Any (or all)
of these combinations could be used for this task. For the purposes of
producing a uniform proper motion survey in the southern hemisphere
to the highest possible accuracy,
we chose the J/R combination for several reasons. Each field has (or soon
will have) a good quality  original photograph on fine--grained emulsion
reaching J~$\sim22.5$, R~$\sim21.0$. The POSS--I `E' and ESO--R material
available are glass copies, are between 0.5~and 1.0~magnitudes less deep,
and neither survey covers the entire southern
hemisphere; the POSS--I plates have the additional complication of not being
on the same system of field centres as the ESO/SERC atlas. The SERC--I survey
lacks depth, reaching I~$\sim19.0$. Moreover, the R~survey was envisaged as
a second epoch survey to the J, and efforts have been made to ensure a good
time baseline between first and second epoch (Morgan et al.~1992;
e.g.\ see Figure~6 later). As we
show in Section~\ref{dcrcorr}, we can accurately model any colour effects
introduced in the astrometry as a result of our choice of J/R plates.
Of course, future enhancements to the SSS may include sophisticated analysis
of the multiple epoch data available from more than two plates per field
if there is sufficient user demand (e.g.\ Paper~{\sc I}).

\begin{figure}
\begin{center}
\setlength{\unitlength}{1mm}
\begin{picture}(80,50)
\includegraphics{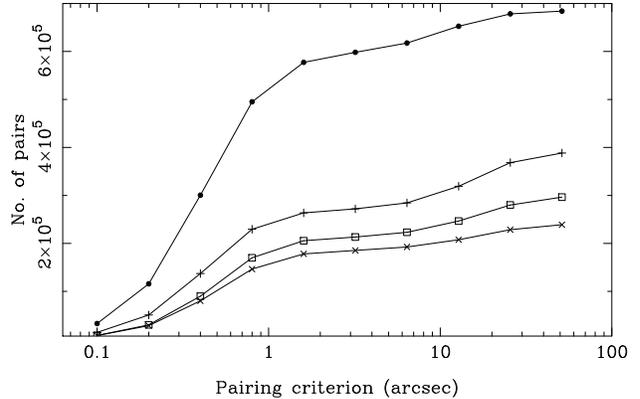}
\end{picture}
\end{center}
\caption[]{The number of objects paired between two plates in fields at
various Galactic latitudes: $b=-32^{\circ}$ (solid circles); $b=-47^{\circ}$
(plus signs); $b=-65^{\circ}$ (open squares) and $b=-87^{\circ}$ (crosses).}
\label{pairing}
\end{figure}

\begin{figure}
\begin{center}
\setlength{\unitlength}{1mm}
\begin{picture}(80,50)
\includegraphics{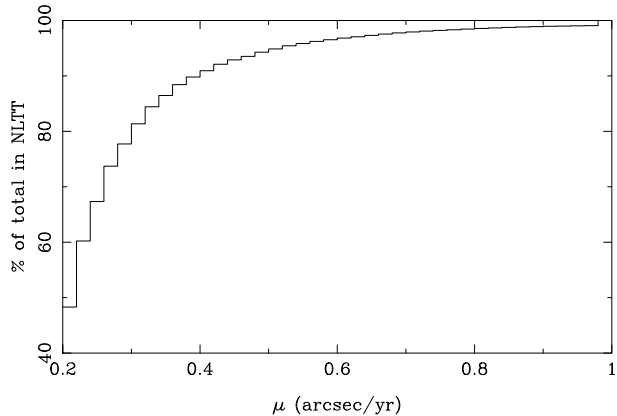}
\end{picture}
\end{center}
\caption[]{Cummulative histogram of the fraction of the total number of
NLTT stars having a proper motion less than $\mu$ arcsec~yr$^{-1}$.
Approximately 90\% of NLTT stars have proper motions less than
$\mu=0.4$ arcsec~yr$^{-1}$.}
\label{highmu}
\end{figure}

Figure~\ref{pairing} shows the number of pairings made between the~J and~R
plates for several different Galactic latitudes. These curves show the same
general trends. Below a pairing criterion of 1~arcsec, a dramatic fall in the
number of pairings occurs due to global positional errors (see 
Section~\ref{results}). For a pairing criterion of more than 6~arcsec, a sudden
and steady rise is seen in the number of pairings at the level of many
thousands at each step. These must be spurious,
since we do not expect $10^4 - 10^5$ high proper motion stars per Schmidt
field. In fact, the {\em total} number of stars having $\mu > 0.2$ 
arcsec~yr$^{-1}$ is of order $10^5$ over the entire sky (Luyten~1979). A
cummulative histogram of the fraction of high proper motion stars
(ie.~those having $\mu > 0.2$ arcsec~yr$^{-1}$) made from the NLTT catalogue
(Figure~\ref{highmu}) shows that 90\% of those catalogued have motions
$\mu < 0.4$ arcsec~yr$^{-1}$. For a median epoch difference of 15~yr
(see Figure~\ref{ephist}), 0.4 arcsec~yr$^{-1}$ corresponds to a shift of
6~arcsec. Hence, in order to pair 90\% of `high' proper motion stars, but at
the same time avoid large numbers of spurious pairings, the pairing
criterion has been set at 6~arcsec. However, the database access software
(see Paper~{\sc I}) allows the user the default option of limiting pairings to 
those within 3~arcsec for most applications when
completeness of high proper motion objects is not
an issue. This is to further minimise the number of spurious pairings for
applications that would be sensitive to such contamination -- clearly, in
Figure~\ref{pairing} even between pairing criteria of~3 and 6~arcsec there is
a rise in pairings over and above that expected from high proper motion objects
alone. As far as pairing motionless objects is concerned, a pairing criterion
of 3~arcsec is very generous in the light of absolute errors in position of
$\sigma_{\alpha,\delta}\sim0.3$~arcsec (Section~\ref{results}).

\subsubsection{Plate--to--plate error mapping}
\label{errmap}

The determination of accurate proper motions requires that systematic positional
errors as a function of plate position be removed from the displacements
of all objects measured between the first and second epoch plates (eg.~EI95).
Such errors arise from emulsion shifts on the photographs themselves, and
also from systematic positional errors during the measurement process. Note that
in Hambly et al.~(1998) we described a procedure for minimising systematic
measurement errors arising from the $xy$ table of SuperCOSMOS; nonetheless we
follow EI95 exactly and compute one--dimensional error functions (ie.~errors
in $x$~and~$y$ as a function of $x$~and~$y$) when mapping the plate--to--plate
systematic errors since small residuals are expected to be present at the few
tenths of a micron level (for example, errors resulting from pixel placement
within the linear CCD used to scan the plate). Other systematic errors possibly
present in the proper motions include those due to positional errors as a 
function of magnitude. For example, it is likely that the centroids of heavily
saturated images off--axis on the photographs are displaced relative to those
of fainter stars owing to off--centre haloes of scattered light, an effect
compounded by the limited dynamic range of the measuring machine. Note
however that such effects cannot be simply measured from the data and removed
since we expect populations of objects having different mean apparent magnitudes
to show systematic differences in mean proper motion due to Galactic structure
(eg.~brighter stars will show systemic motions with respect to fainter stars
since they are, on average, nearer to the Sun). We argue that such effects will
be minimised by using the photographic material specified in the
previous Section since the first and
second epoch plates were obtained on the same field centres with the same
telescope using the same guide stars at similar times of the year and at similar
(optimum) airmass. Other effects due to misalignment of the scanning beam with
the plate along with different `depths' of image within the emulsion 
are likely to be present -- eg.~EI95. In their work such
effects were eliminated by measuring each plate
twice, at orientations within the machine of $0^{\circ}$ and~$180^{\circ}$.
Taking the mean position between the two scans, these errors were removed. 
The scale of the SSS makes scanning each plate twice impractical and so 
systematic errors resulting from such effects will remain in these data.

\subsubsection{Zeropoint of the proper motions}

\begin{figure*}
\begin{center}
\setlength{\unitlength}{1mm}
\begin{picture}(150,90)
\includegraphics{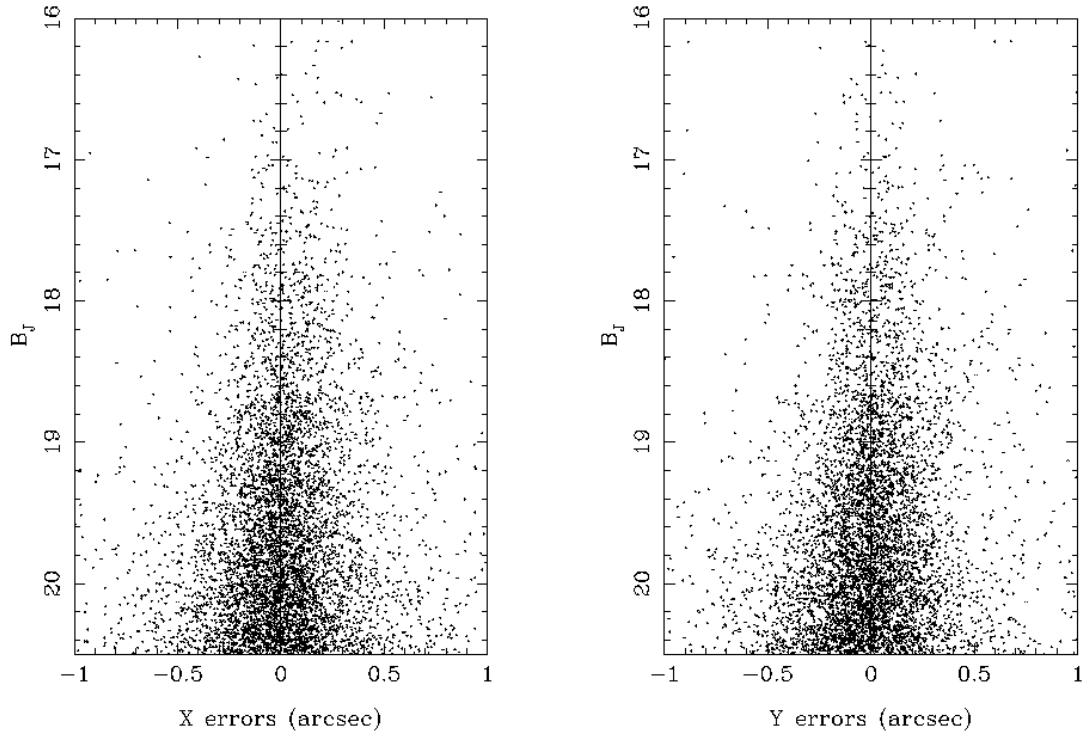}
\end{picture}
\end{center}
\caption[]{Distribution of residual displacement in galaxian images in $xy$
as a function of magnitude for the SGP field~411, after plate--to--plate
error mapping and zeropoint translation between first and second epoch
plates.}
\label{galzero}
\end{figure*}

\begin{figure*}
\begin{center}
\setlength{\unitlength}{1mm}
\begin{picture}(150,90)
\includegraphics{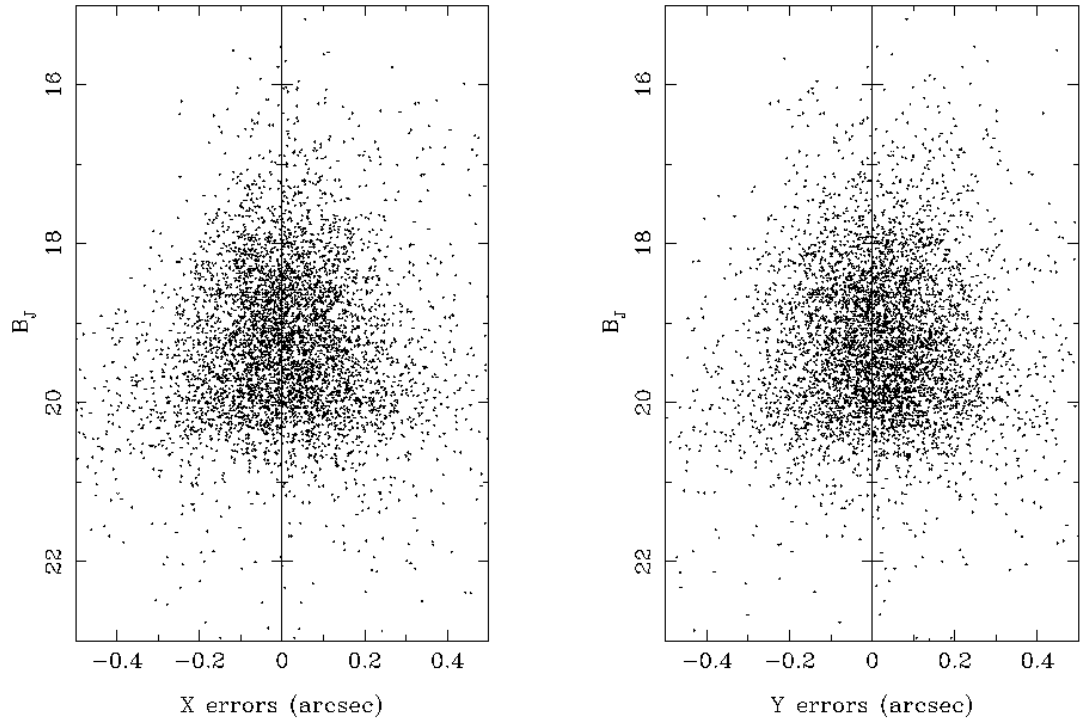}
\end{picture}
\end{center}
\caption[]{Distribution of residual displacement (between first and
second epoch plates) in QSO images in $xy$
as a function of magnitude for the entire SGC survey (objects catalogued
in Veron--Cetty \& Veron~1998). Median zeropoints and RMS scatter as a
function of magnitude are tabulated in Table~\ref{qsonum}.}
\label{qso}
\end{figure*}

\begin{figure}
\begin{center}
\setlength{\unitlength}{1mm}
\begin{picture}(80,80)
\includegraphics{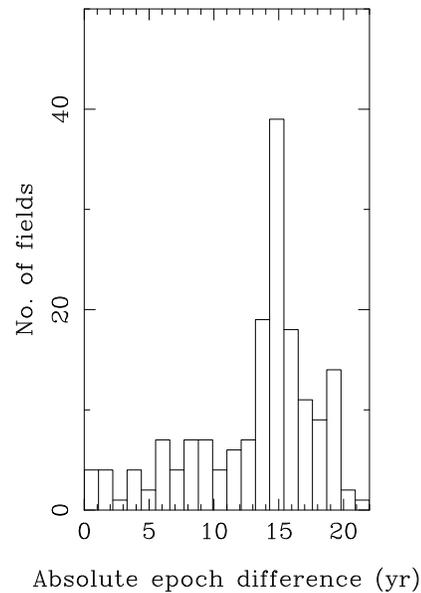}
\end{picture}
\end{center}
\caption[]{Histogram of epoch differences between the SGC~B$_{\rm J}$ 
and~R photographic observations.}
\label{ephist}
\end{figure}

The zeropoint of the proper motion system is fixed using objects classed
as galaxies (Paper~{\sc II} describes the image classification procedure).
As described in EI95, once the plate--to--plate error mapper
has removed systematic errors between the first and second epoch plates, in
general the mean stellar displacement will be zero (since stellar images
dominate the number counts) while that of galaxies will be non--zero. Simply
applying a global translation to make the galaxy images have zero mean
displacement fixes the proper motion zeropoint to the extragalactic frame.
Note that the image classifier is reliable at the level of $\geq90$\% down
to B$_{\rm J}\sim20.5$ (Paper~{\sc II}, Table~9) 
and so only galaxies brighter than that
magnitude are used in this procedure. Figure~\ref{galzero} shows an example
of the distribution in galaxy displacement as a function of magnitude for
the SGP field~411, demonstrating the effectiveness of this procedure.
As an independent and external check on the extragalactic zeropoint, we also
examined the displacements of all isolated, good quality images paired between
the SGC survey database and the Veron--Cetty \& Veron~(1998) 
QSO catalogue. Figure~\ref{qso} shows this comparison ($>5000$~QSOs over the
entire SGC region).
Table~\ref{qsonum} details the mean displacements and RMS errors for the QSO
sample. Clearly, there are no significant non--zero residuals. Moreover, this
test illustrates the likely level of errors in the stellar proper motions.
Figure~\ref{ephist} shows a histogram of the epoch differences between the~J
and~R survey photographs within the SGC survey. For a median epoch difference
of 15~yr, the final column in Table~\ref{qsonum} gives the implied
mean proper motion precision and RMS zero point error, in either co-ordinate,
as a function of magnitude. Obviously, some fields have much shorter
time baselines than the median value and the proper motions in such fields
will be correspondingly less accurate (see Section~\ref{pmerrs}).
\begin{center}
\begin{table*}
\begin{tabular}{crrcc}
Magnitude            &\multicolumn{2}{c}{Zeropoint and RMS (arcsec)} & 
RMS zeropoint & $\sigma_{\mu_{\alpha,\delta}}^1$ \\
  range              &\multicolumn{1}{c}{RA}&\multicolumn{1}{c}{DEC} & 
error$^1$ (mas~yr$^{-1})$ & (mas~yr$^{-1}$)\\
\multicolumn{5}{c}{ }\\
 $15.5<{\rm B}_{\rm J}<16.5$ & $ 0.015\pm0.132$ & $ 0.066\pm0.225$ &   2.2 &  11.9 \\
 $16.5<{\rm B}_{\rm J}<17.5$ & $ 0.043\pm0.137$ & $ 0.052\pm0.148$ &   2.2 &   9.5 \\
 $17.5<{\rm B}_{\rm J}<18.5$ & $ 0.007\pm0.127$ & $ 0.024\pm0.133$ &   0.8 &   8.7 \\
 $18.5<{\rm B}_{\rm J}<19.5$ & $ 0.004\pm0.132$ & $ 0.019\pm0.133$ &   0.7 &   8.9 \\
 $19.5<{\rm B}_{\rm J}<20.5$ & $ 0.005\pm0.164$ & $ 0.027\pm0.152$ &   0.9 &  10.5 \\
 $20.5<{\rm B}_{\rm J}<21.5$ & $ 0.002\pm0.242$ & $ 0.016\pm0.214$ &   0.5 &  15.2 \\
 $21.5<{\rm B}_{\rm J}<22.5$ & $ 0.044\pm0.395$ & $-0.066\pm0.435$ &   2.6 &  27.7 \\
\multicolumn{5}{c}{ }\\
\multicolumn{5}{l}{$^1$Assuming median epoch difference of 15~yr.}\\
\end{tabular}
\caption[ ]{Residual systematic zeropoint errors in QSO positions between
first and second epoch plates from
Figure~\ref{qso}, and implied proper motion errors (both random and systematic)
for an epoch baseline of 15~yr, as a function of magnitude (however see
Section~\ref{spmerr}).}
\label{qsonum}
\end{table*}
\end{center}

\subsubsection{Correction of systematic effects due to differential colour
refraction (DCR)}
\label{dcrcorr}

\begin{figure}
\begin{center}
\setlength{\unitlength}{1mm}
\begin{picture}(80,120)
\includegraphics{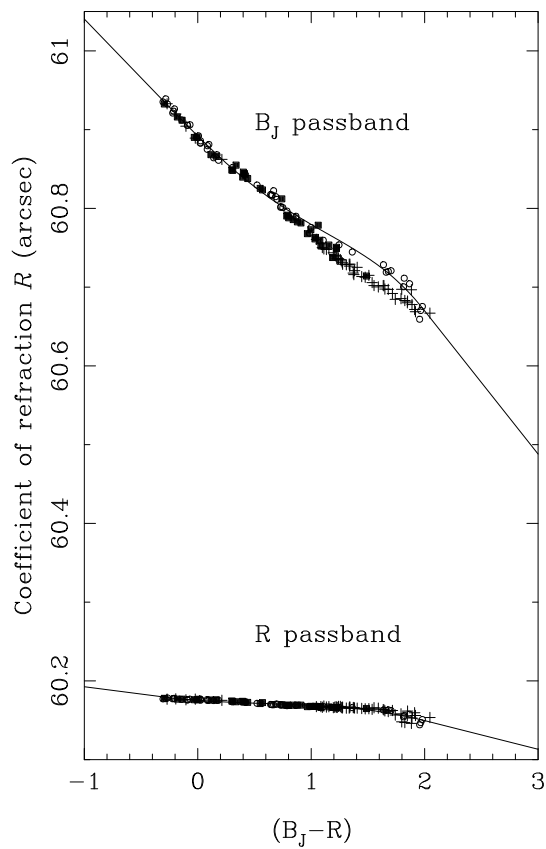}
\end{picture}
\end{center}
\caption[]{Computational models of $R$, the coefficient of refraction, as
a function of (B$_{\rm J}$--R) colour. The symbol types are + for giants, filled
squares for subgiants and open circles for dwarfs. The solid lines are
least--squares polynomial fits to the dwarf data along with linear extrapolations
to provide an estimate of $R$ for any colour (cf.~Evans \& Irwin~1995,
Figures~11 and~12).}
\label{dcr}
\end{figure}

As noted in EI95 differential colour refraction (hereafter DCR) corrections
will be required in order that objects of different colour to the average
have no systematic error in proper motions measured from plates of different
bandpass. Again, these errors must not be measured from the data since 
populations of different mean colour (eg.~redder young disk stars having
relatively high metallicity and bluer old halo stars having relatively low
metallicity) are {\em expected} to have different systemic motions due to
their different ensemble kinematics. We modelled DCR effects following exactly
the prescription given in EI95 with the necessary changes made for the different
emulsion/filter combinations used here, with relevant data being taken from
Evans~(1989) and references therein. In Figure~\ref{dcr} we show our 
computations of the coefficient of refraction $R$ as a function of synthetic
(B$_{\rm J}$--R) colour for the~B$_{\rm J}$ and~R passbands along with the polynomial fits and
linear extrapolations used.

The DCR corrections are applied as detailed in Equations~1 and~2, Section~5
of EI95. The corrections take the form of small changes to the second
epoch positions as a function of colour and as a function of mean galaxy
colour (since the galaxies define the zeropoint described above).
Note that in computing the refraction coefficient we assumed nominal
atmospheric temperature, pressure, humidity and unit airmass. Calculation
shows that this is good enough for our purposes -- for example, recomputing
$R$~for airmass~$\chi$
in the range $1.0\leq\chi\leq2.0$ in steps of~0.1 shows
negligible zeropoint changes in $R$ as a function of colour while the gradient
remains constant.

\subsubsection{Proper motion error estimates}
\label{pmerrs}

In order that some estimate of probable errors in proper 
motion are available when
using SSS data, error values in units of mas~yr$^{-1}$ are provided with each
proper motion. These errors are estimated on a field--by--field basis, and 
therefore take into account the small differences in survey plate quality as
well as the varying time baselines. The estimates also take into account the
effects of centroiding precision as a function of magnitude, individually for
stars and galaxies. The measurements are made from the global average
dispersions in centroids, as a function of magnitude, once the error 
mapping described in Section~\ref{errmap} has been completed. 
Of course, the assumption in making these
estimates is that the intrinsic proper motion is small compared to the errors
and makes negligible contribution to the overall dispersion.
This is obviously true for galaxies and faint stars, but is an increasingly poor
assumption for increasingly brighter stars. The errors are thus overestimated 
for bright stars.

\section{Results and discussion}
\label{results}

\subsection{Positions}

\subsubsection{Residuals of the fitting standards}

\begin{figure*}
\begin{center}
\setlength{\unitlength}{1mm}
\begin{picture}(150,90)
\includegraphics{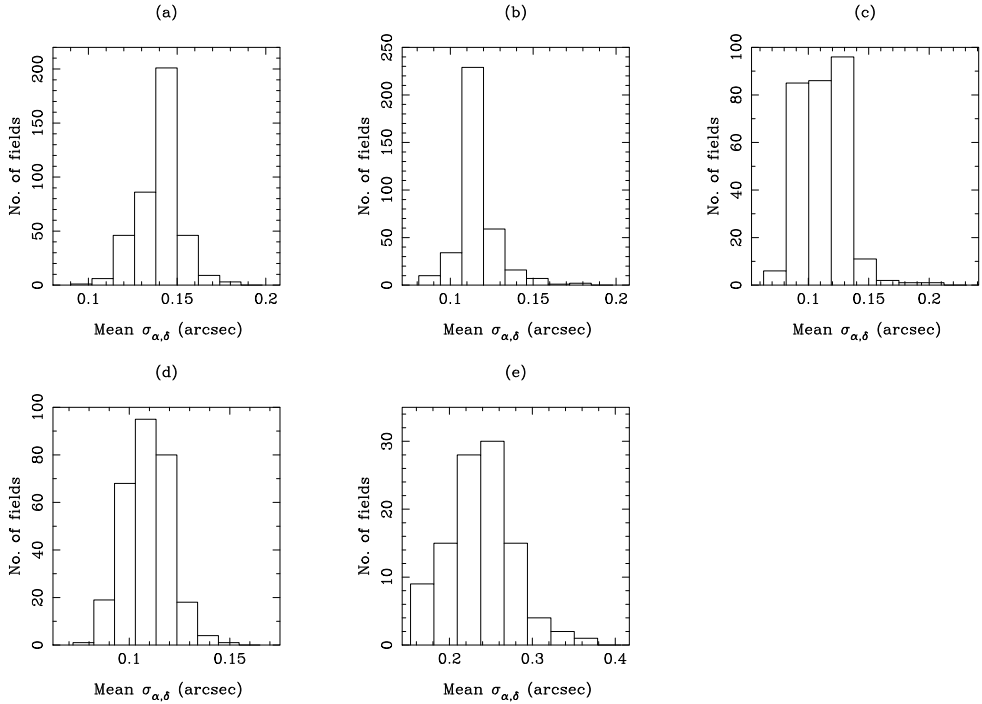}
\end{picture}
\end{center}
\caption[]{Histograms of the mean RMS residual per star per SGC plate in either
co-ordinate for reference catalogue (ie.~Tycho--2) stars (a) SERC~J/EJ,
(b) SERC ER/AAO--R (excluding data from 4415 film), (c) SERC--I, 
(d) ESO--R and (e) POSS--I E.}
\label{tychores}
\end{figure*}

Figure~\ref{tychores} shows histograms of the mean RMS error per star per
SGC plate in Right Ascension and Declination
for the different survey plate collections included in
the SSS survey. These plots demonstrate the ability of the plate solutions to 
predict the positions of the reference catalogue stars
{\em on the plates}. It is encouraging
that even for the 51~POSS--I E~glass copies -- which are not of the highest
quality compared to modern atlas copies, and which
have a mean epoch of~1954 (Minkowski \& Abell~1963) -- the global plate
solutions are accurate at the 0.3~arcsec level.

The modal values from these histograms are in the range $\sim0.1$ to
$\sim0.15$~arcsec, depending on survey type and excluding the POSS--I E data.
The SERC--I survey gives the best results, with some mean residuals down at 
$\sim60$~mas. This result is probably
due to a combination of minimal proper motion errors (the mean observational
epoch of the SERC--I survey is close to that of Hipparcos/Tycho) and also the
fact that the I~plates are the least sensitive of the surveys, resulting in less
saturation and lower intensity scattered light haloes for the reference catalogue
standards.

\subsubsection{Empirical uncertainty in positions relative to the International
Celestial Reference Frame (ICRF)}
\label{absacc}

In a recent paper, Deutsch~(1999) compared positions of a sample of extragalactic
radio sources with those derived from several currently available wide--field
digitised photographic survey databases. The comparison sample came from the
very long baseline interferometric study of Ma et al.~(1998), including
212~sources {\em defining} the ICRF. As in Deutsch~(1999), we included
all defining, candidate and `other' sources from Ma et al., since their
positional uncertainties are orders of magnitude smaller than those 
expected from
photographic data. Within the SSS survey object catalogue
database, we matched up sources within
3~arcsec of the quoted ICRF position but excluded poor quality images 
(e.g.~those near very bright stars, step--wedges or plate labels), images 
below Galactic latitude $|b|=30^{\circ}$ (since we are concerned with deep,
sky limited plates) and deblended images
(whose centroids will be, in general, systematically wrong -- eg.~Beard,
MacGillivray \& Thanisch~1990). Table~\ref{icrf} shows the results of the
comparison, where we have split the sample into stars (for `stars' in this
discussion we are referring to extragalactic objects classed as stellar in
SSS catalogues -- presumably these are QSOs) and galaxies since
image centroiding is in general a factor $\sim2$ less accurate 
for galaxies than 
for stars (eg.~Irwin~1985). We also provide figures for stellar samples 
having magnitudes more than 2.5$^{\rm m}$ above the respective plate limits
where the random errors due to emulsion noise are reasonably small and
constant (eg.~Lee \& van Altena~1983). We give results equivalent to those
in Deutsch~(1999), for each of the survey types, to enable direct comparison:
columns~3 and~4 of Table~\ref{icrf} show the median zeropoint of the respective
samples (described by columns~1 and~2) with respect to the ICRF; column~5 gives
the number of objects in each sample; columns~6 through to~14 are grouped in
triplets, each set of three showing the maximum deviation in $\alpha,\delta$ and
R($=\{\alpha^2+\delta^2\}^{1/2}$) for the best 68\%, 95\% and 99\% of the samples.
\begin{center}
\begin{table*}
\begin{tabular}{lcrrcccccccccc}
Survey & Object & $<\Delta\alpha>$&$<\Delta\delta>$& N & 
$\Delta\alpha$ & $\Delta\delta$ & $\Delta{\rm R}$ &
$\Delta\alpha$ & $\Delta\delta$ & $\Delta{\rm R}$ &
$\Delta\alpha$ & $\Delta\delta$ & $\Delta{\rm R}$ \\
       &  type  & \multicolumn{2}{c}{(arcsec)}  &   & 
\multicolumn{3}{c}{68\% (arcsec)}& \multicolumn{3}{c}{95\% (arcsec)}& 
\multicolumn{3}{c}{99\% (arcsec)}\\ 
\multicolumn{14}{c}{ }\\
% 
%  For Tycho-2:
%
SERC--J/ & All stars & 
 $ -0.054$&$ -0.041$&  110 & 0.120 & 0.117 & 0.167 & 0.319 & 0.307 & 0.443 & 0.637 & 0.649 & 0.909 \\
 EJ          & All galaxies &
 $ -0.100$&$  0.328$&    5 & 0.145 & 0.456 & 0.478 & 0.502 & 0.880 & 1.013 & 0.502 & 0.880 & 1.013 \\
           & Stars, B$_{\rm J}<20.0$ &
 $ -0.022$&$ -0.019$&   92 & 0.108 & 0.108 & 0.152 & 0.238 & 0.241 & 0.338 & 0.313 & 0.317 & 0.445 \\
\multicolumn{14}{c}{ }\\
SERC--ER/ & All stars & 
 $ -0.093$&$  0.069$&  103 & 0.144 & 0.229 & 0.270 & 0.424 & 0.512 & 0.665 & 0.641 & 0.686 & 0.939 \\
AAO--R     & All galaxies &
 $ -0.117$&$  0.187$&    4 & 0.507 & 0.364 & 0.625 & 0.628 & 0.932 & 1.124 & 0.628 & 0.932 & 1.124 \\
           & Stars, R~$<18.5$ &
 $ -0.072$&$  0.079$&   82 & 0.152 & 0.224 & 0.271 & 0.278 & 0.405 & 0.491 & 0.402 & 0.502 & 0.644 \\
\multicolumn{14}{c}{ }\\
SERC--I & All stars & 
 $ -0.160$&$  0.081$&   54 & 0.224 & 0.312 & 0.384 & 0.565 & 0.554 & 0.791 & 1.098 & 1.339 & 1.732 \\
           & All galaxies &
 $ -0.135$&$ -0.304$&    7 & 0.170 & 0.251 & 0.303 & 0.462 & 0.631 & 0.782 & 0.462 & 0.631 & 0.782 \\
           & Stars, I~$<17.0$ &
 $ -0.110$&$  0.113$&   21 & 0.121 & 0.284 & 0.309 & 0.324 & 0.474 & 0.575 & 0.347 & 0.548 & 0.648 \\
\multicolumn{14}{c}{ }\\
ESO--R & All stars & 
 $  0.083$&$  0.135$&   38 & 0.133 & 0.173 & 0.218 & 0.283 & 0.316 & 0.424 & 0.450 & 0.398 & 0.600 \\
           & All galaxies &
 $  0.365$&$  0.381$&    2 & 0.251 & 0.235 & 0.344 & 0.251 & 0.267 & 0.367 & 0.251 & 0.267 & 0.367 \\
           & Stars, R~$<18.0$ &
 $  0.096$&$  0.146$&   23 & 0.134 & 0.178 & 0.223 & 0.355 & 0.305 & 0.468 & 0.437 & 0.386 & 0.583 \\
\multicolumn{14}{c}{ }\\
POSS--I E & All stars & 
 $  0.130$&$ -0.059$&   22 & 0.328 & 0.312 & 0.453 & 1.235 & 1.336 & 1.819 & 1.563 & 1.752 & 2.348 \\
           & Stars, E~$<17.5$ &
 $  0.172$&$ -0.005$&    6 & 0.141 & 0.194 & 0.240 & 0.333 & 0.510 & 0.609 & 0.333 & 0.510 & 0.609 \\
\multicolumn{14}{c}{ }\\
\end{tabular}
\caption[ ]{Empirical uncertainty estimates in $|b|>30^{\circ}$
celestial positions relative
to the ICRF (after Deutsch~1999). For these purposes, `stars' refers to 
point--like extragalactic objects, presumably QSOs.}
\label{icrf}
\end{table*}
\end{center}

Comparing the numbers in Table~\ref{icrf} with corresponding values in Table~1
of Deutsch~(1999), the accuracy of the SGC survey global astrometry is clearly
as good as any other database for the same survey material. Table~\ref{icrf}
also clearly demonstrates that, globally at least, the astrometry of
point--like sources has no 
zeropoint errors with respect to the ICRF larger than $\sim100$~mas for many
of the plate collections. In the worst cases (e.g.~POSS--I~E, ESO--R)
the global zeropoint errors are no larger than $\sim200$~mas. Note that these
comparison objects have magnitudes in the range $15<{\rm R}<19$ 
(e.g.~Deutsch~1999 Figure~1d) while the reference catalogue stars 
have R~$<12$. Note also that some proportion
of the apparently random errors listed in Table~\ref{icrf} will manifest
themselves as local, systematic position errors (see the next Section).

\subsubsection{Local position errors as a function of magnitude and plate 
position}

As stated earlier, it is important to investigate systematic errors as a 
function of magnitude and plate position since the external reference
catalogue standards are highly saturated on the survey photographs. Off--centre
scattered light haloes inevitably `pull' the centroids of these stars (as
measured by the scanning machine) in a radial direction from the plate centre.
The direction of this shift on the celestial sphere will be discontinuous as
one moves from using positional data in one field to that in an adjacent field;
the amplitude of this discontinuity will also vary with magnitude. The
discontinuities in systematic errors with respect to external data are
likely to be highest in fainter magnitude ranges for global astrometric fits
based on bright standards.

\begin{figure*}
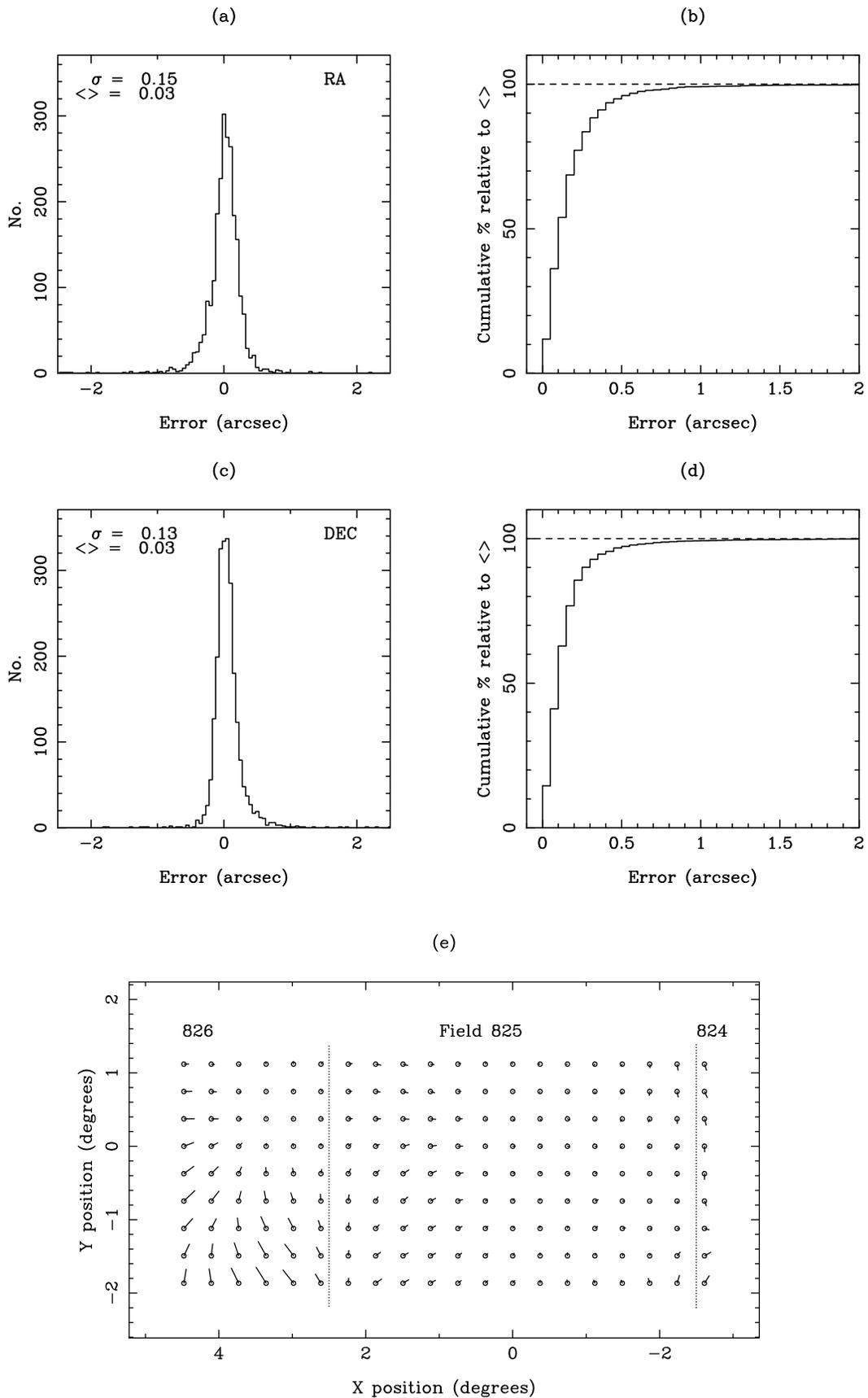

\begin{center}
\setlength{\unitlength}{1mm}
\begin{picture}(170,220)
\includegraphics{figure9.eps}
\includegraphics{figure9e.eps}
\end{picture}
\end{center}
\caption[]{Comparison against SDSS CCD calibration region `A' data 
(field centre $\alpha\sim0^{\rm h}40^{\rm m}$, $\delta\sim0^{\circ}$) from
Stone et al.~(1999) in the magnitude range $10<{\rm R}<14$. In (e), the
systematic error vectors have been scaled such that 1~arcsec~$\equiv1^{\circ}$.
Dotted lines indicate the field data boundaries.}
\label{stone1014}
\end{figure*}

\begin{figure*}
\begin{center}
\setlength{\unitlength}{1mm}
\begin{picture}(170,220)
\includegraphics{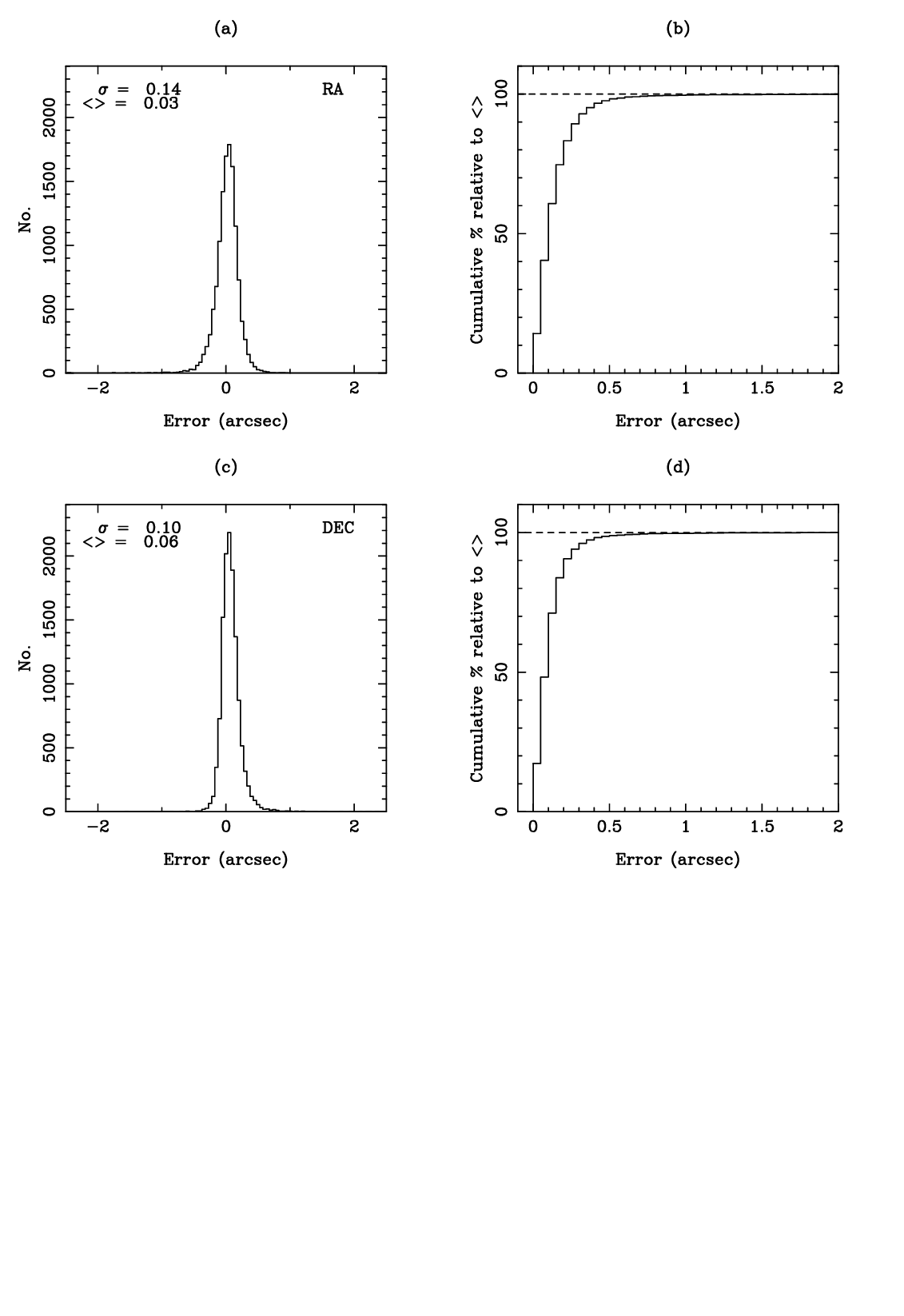}
\includegraphics{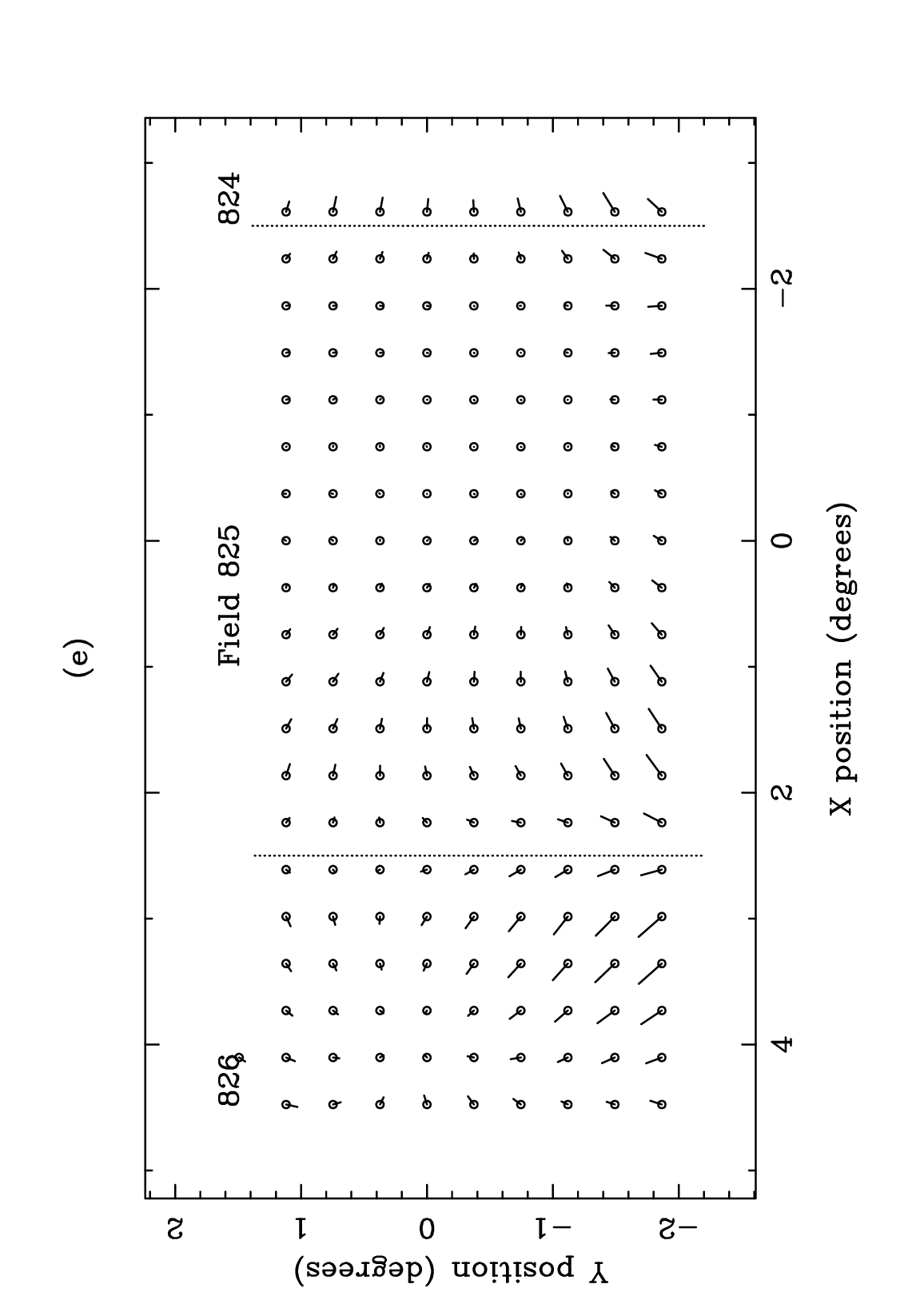}
\end{picture}
\end{center}
\caption[]{Comparison against SDSS CCD calibration region `A' data
(field centre $\alpha\sim0^{\rm h}40^{\rm m}$, $\delta\sim0^{\circ}$) from
Stone et al.~(1999) in the magnitude range $14<{\rm R}<18$. In (e), the
systematic error vectors have been scaled such that 1~arcsec~$\equiv1^{\circ}$.
Dotted lines indicate the field data boundaries.}
\label{stone1418}
\end{figure*}

Stone, Pier \& Monet~(1999) provide the means to measure the size of such
systematic effects via comparison with their CCD drift--scan data in the
Sloan Digital Sky Survey calibration regions. These regions consist of equatorial 
strips of $\sim7.5^{\circ}$ in Right Ascension and 
$\sim3.5^{\circ}$ in Declination with typical
astrometric accuracy quoted as $\sim35$~mas at R=10.0 degrading 
to $\sim70$~mas
at R=18.0. Figures~\ref{stone1014} and~\ref{stone1418} show the results of
this comparison for the R~data in the range $10<{\rm R}<14$ and 
$14<{\rm R}<18$
respectively, for calibration region `A' which falls over survey 
fields~824 to~826. The extent of the individual field data are indicated in the
plots, where a `seamless' catalogue has been extracted from the database as
detailed in Paper~{\sc I}. Data from the $>10000$ objects in common have been
binned up in boxes $\sim20$~arcmin on a side and smoothed/filtered on a 
$3\times3$ bin scale for the purposes of display in Figures~\ref{stone1014}(e) 
and~\ref{stone1418}(e).


Once again, we find no large systematic offset between the SGC co-ordinates and 
those of Stone et al.~(1999). The random errors are also consistent with those
from Table~\ref{icrf}, and are estimated from the distributions in 
Figures~\ref{stone1014},\ref{stone1418}~(a) and~(c) to be in the range
$0.10<\sigma_{\alpha,\delta}<0.16$ arcsec. 
Figures~\ref{stone1014},\ref{stone1418}~(e) demonstrate the level of systematic
differences as a function of magnitude (error vectors are scaled such that
$1^{\circ}\equiv1$~arcsec). A discontinuity in the vector direction can clearly
be seen in Figure~\ref{stone1418}(e) when moving from field~826 to~825; the
level of this discontinuity is several tenths of an arcsec. In any application 
having a tight error budget (eg.~fibre spectroscopy with very small diameter
fibres, say 0.5~arcsec or smaller) one should be careful when using these
data across a field boundary. We note, however, that for 2dF applications
(ie.~2~arcsec diameter fibres over $2.0^{\circ}$, Gray et al.~1993)
these data have the requisite astrometric precision.

The residuals seen in Figure~\ref{stone1014}(e) at X~$\sim3.0^{\circ}$, 
Y~$\sim-1.5^{\circ}$ illustrate how potentially misleading the histograms in
Figure~\ref{tychores} can be for estimating true external errors in the
celestial co-ordinates. Clearly, there are systematic discrepancies in SGC
survey astrometry on scales of a degree or so and at the level of up to
$\sim0.3$~arcsec that are not apparent from the residuals of the individual
field astrometric fits. We reiterate that Figure~\ref{tychores} merely shows
how well the astrometric fits model how the reference stars appear on the 
plate to the measuring machine, and 
{\em not necessarily} where they are on the celestial sphere.

The standard errors of $\sigma_{\alpha,\delta}\sim0.1$~arcsec in 
Figures~\ref{stone1014},\ref{stone1418}~(a) and~(c) are dominated by those for
the most numerous, faint stars at R~$\sim18.5$. Towards the plate limits,
object centroiding becomes less accurate due to emulsion noise (eg.~Lee \&
van Altena~1983) and we expect that the errors in positions will increase
with magnitude (eg.~Hambly et al.~1998, Figure~8) such that
at the plate limits of R~$\sim21$, the random errors will be
$\sigma_{\alpha,\delta}\sim0.3$~arcsec.

\subsection{Proper motions}
\label{spmerr}

Platais et al.~(1998) describe the Yale/San~Juan SPM programme, from which
the first release of data contains proper motions of 58,880 objects (defined
by a set of input catalogues) around the south Galactic pole (SGP). The
typical accuracies of their data range from $\sim3$~mas~yr$^{-1}$ at V~$<15$
to $\sim8$~mas~yr$^{-1}$ at V=$18.5$.

\begin{figure*}
\begin{center}
\setlength{\unitlength}{1mm}
\begin{picture}(100,220)
\includegraphics{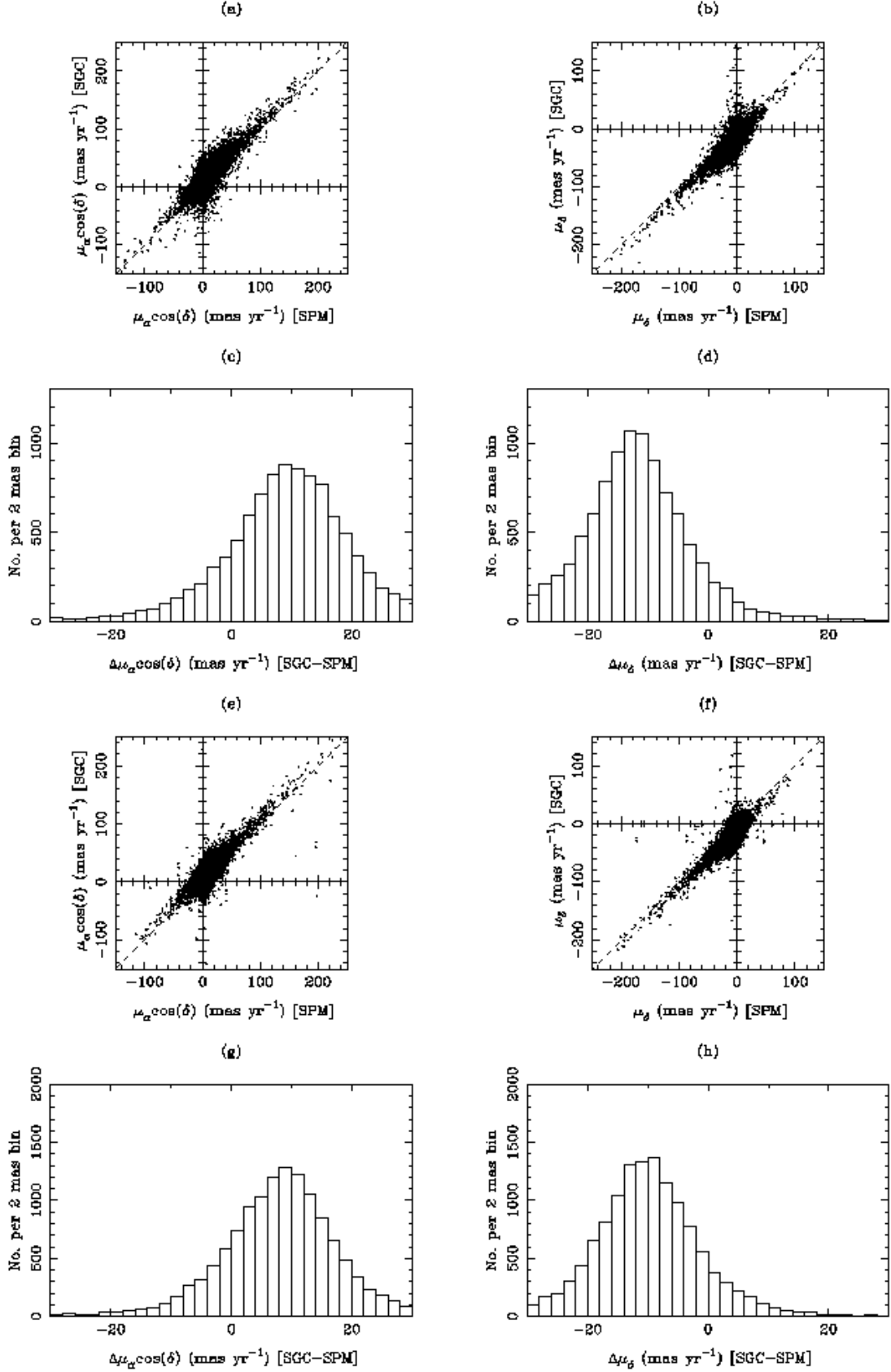}
\end{picture}
\end{center}
\caption[]{Comparison of SGC proper motions versus those from SPM;
(a) to (d) are for $10.0<{\rm V}<14.0$, (e) to (h) for $14.0<{\rm V}<16.0$}
\label{spm1}
\end{figure*}

\begin{figure*}
\begin{center}
\setlength{\unitlength}{1mm}
\begin{picture}(100,220)
\includegraphics{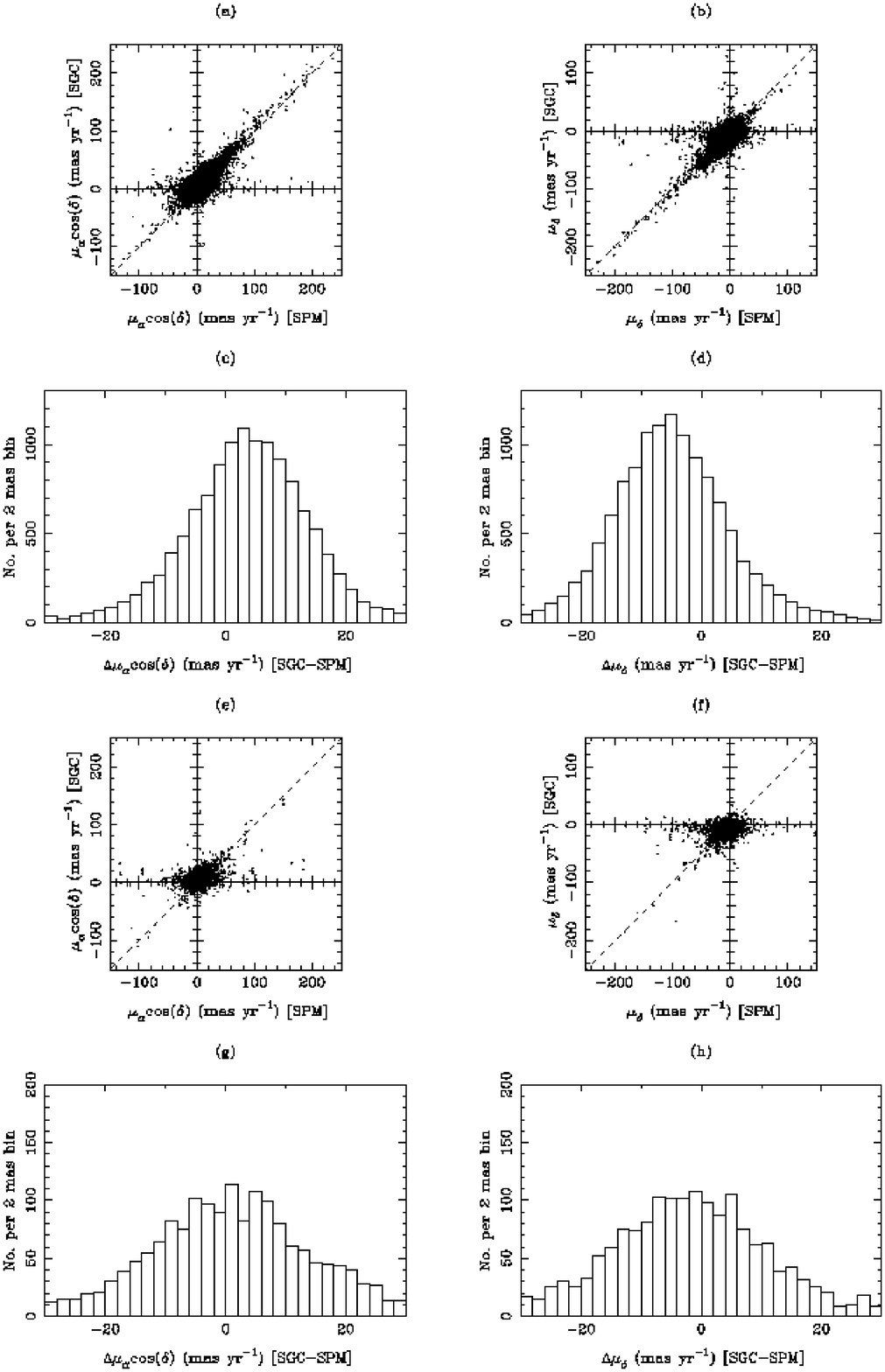}
\end{picture}
\end{center}
\caption[]{Comparison of SGC proper motions versus those from SPM;
(a) to (d) are for $16.0<{\rm V}<17.5$, (e) to (h) for $17.5<{\rm V}<18.5$}
\label{spm2}
\end{figure*}

Figures~\ref{spm1} and~\ref{spm2} show straight--forward comparisons between
SPM and SGC proper motions in four magnitude ranges: $10<{\rm V}\leq14$,
$14<{\rm V}\leq16$, $16<{\rm V}\leq17.5$ and $17.5<{\rm V}\leq18.5$ (the latter
is limited by the SPM data). It is immediately apparent that there are magnitude
dependent systematic errors in the SGC proper motions of brighter stars.
Table~\ref{spmtab} quantifies these zeropoint errors for the SGP region and 
also gives the scatter between the SGC and SPM data. Given the level
of random error quoted in Platais et al.~(1998) these values should be
dominated by errors in the SGC measurements.
\begin{center}
\begin{table*}
\begin{tabular}{crrrrcc}
Magnitude & \multicolumn{2}{c}{$\mu_{\alpha}\cos(\delta)$ zp }
& \multicolumn{2}{c}{$\mu_{\delta}$ zp and $\sigma$} & 
$\sigma_{\mu_{\alpha}\cos(\delta)}^1$ & $\sigma_{\mu_{\delta}}^1$\\
  range   &  \multicolumn{2}{c}{and $\sigma$ (mas~yr$^{-1}$)} &  \multicolumn{2}{c}{(mas~yr$^{-1}$)} &  \multicolumn{2}{c}{(mas~yr$^{-1}$)} \\ 
\multicolumn{7}{c}{ }\\
 10.0~$<{\rm V}<$~14.0&$  10.0$&$   9.5$&$ -12.4$&$   8.2$&$  16.4$&$  18.9$\\
 14.0~$<{\rm V}<$~16.0&$   8.5$&$   8.5$&$ -10.3$&$   7.8$&$  13.8$&$  15.9$\\
 16.0~$<{\rm V}<$~17.5&$   3.9$&$   9.6$&$  -5.4$&$   9.1$&$  10.7$&$  11.3$\\
 17.5~$<{\rm V}<$~18.5&$   0.8$&$  14.3$&$  -1.8$&$  14.1$&$  14.2$&$  14.3$\\
\multicolumn{7}{c}{ }\\
\multicolumn{7}{l}{$^1$`Effective' global random error including 
zeropoint (zp) error.}\\
\end{tabular}
\caption[ ]{Quantitative comparison between SGC and SPM proper motions
(see text).}
\label{spmtab}
\end{table*}
\end{center}
Comparing with Figure~\ref{qso} and Table~\ref{qsonum} there is a consistent
picture of no magnitude systematics for m~$\geq17.5$. However, brighter
magnitude ranges in Table~\ref{qsonum} fail to indicate true zeropoint errors.
This underestimate
is due to the fact that the data cover the whole SGC survey region, and systematic
zeropoints of different magnitude and sign manifest themselves as an increased
random error. On the other hand, because the SGC--SPM comparison is done in a
restricted declination range near the SGP, there is a clearer tendency for
systematic zeropoints of the same magnitude and sign. The 
global random errors for SGC proper motions are therefore dominated by
magnitude dependent systematics for V~$<17$, and Table~\ref{spmtab} indicates
this by measuring the apparent SGC--SPM scatter without removal of zeropoint
errors. Application of a global correction to the proper motions to remove 
these zeropoint errors is clearly not possible since their magnitude and sign
changes over the SGC region. Once again,
centroiding errors will increase with magnitude as one approaches the
plate limits such that at J,R~$\sim22,21$ the proper motion errors will be 
$\sim50$~mas~yr$^{-1}$, but with zeropoint error with respect
to the extragalactic frame of $\leq1$~mas~yr$^{-1}$. 

\section{Conclusion}
\label{concs}

We have presented a detailed description of the astrometric properties of the
SuperCOSMOS Sky Survey (SSS)\footnote{database available
online at \verb+http://www-wfau.roe.ac.uk/sss+}. Using the first release
South Galactic Cap (SGC) data, we have demonstrated the
level of random errors in positions and proper motions by comparison 
SGC survey with
external data; moreover we have demonstrated the level of any systematic
effects in these astrometric parameters as a function of magnitude, and
in the case of positions as a function of field position. For a summary of
these results along with similar comparisons for other large--scale survey
programmes and a guide to using the SSS database, the reader is
referred to Paper~{\sc I}.

\section*{Acknowledgements}

We thank Ron Stone for providing the SDSS data in advance of publication and in
machine--readable form. We would also like to thank Russell Cannon,
Sue Tritton, Mike Read and
Patrick Wallace for useful discussions. NCH thanks Peter Draper for advice
concerning FITS WCS keywords and the use of GAIA. Funding for the 
University of Edinburgh Institute
for Astronomy Wide--Field Astronomy Unit and Institute of Astronomy Cambridge
Astronomical Survey Unit is provided by the UK PPARC. This research has made
use of data archived at the CDS, Strasbourg. We acknowledge the use of 
{\sc Starlink} computer facilities at Edinburgh and Leicester.
We are indebted to the referee, Sean Urban, for a prompt and 
thorough review of these manuscripts.

The National Geographic Society--Palomar Observatory Sky Survey (POSS--I) was
made by the California Institute of Technology with grants from the National
Geographic Society. The UK Schmidt Telescope was operated by the Royal
Observatory Edinburgh, with funding from the UK Science and Engineering 
Research Council (later the UK Particle Physics and Astronomy Research Council),
until 1988~June, and thereafter by the Anglo--Australian Observatory. The blue
plates of the Southern Sky Atlas and its equatorial extension (together known
as the SERC--J/EJ) as well as the Equatorial Red (ER), the second epoch
(red) Survey (SES or AAO--R) and the infrared (SERC--I) Survey
were taken with the UK Schmidt Telescope. All data retrieved from URLs
described herein
are subject to the copyright given in this copyright summary. Copyright
information specific to individual plates is provided in the downloaded FITS
headers.

\vfill
\bsp

\end{document}